\documentclass[multphys,vecphys]{svmult}

\usepackage{comment}
\usepackage{amsmath}

\usepackage{hyperref}
\hypersetup{colorlinks=true,
            citecolor=blue,
            linkcolor=blue}

\def\ez {z}
\def\eb {\bar z}

\makeindex             

\begin{document}

\title*{Complete determination of the orbital parameters of a system with $N+1$
bodies using a simple Fourier analysis of the data}

\titlerunning{Orbital parameters with a Fourier analysis} 

\author{Alexandre C. M. Correia}
\authorrunning{A.C.M. Correia} 
\institute{CFisUC, Dep. F\'isica, Univ. Coimbra, 3004-516 Coimbra, Portugal. 
\\
IMCCE, 
PSL Universit\'e, 77 Av. Denfert-Rochereau, 75014 Paris, France}


\maketitle

\begin{abstract}
Here we show how to determine the orbital parameters of a system composed of a
star and $N$ companions (that can be planets, brown-dwarfs or other stars), using a
simple Fourier analysis of the radial velocity data of the star.
This method supposes that all objects in the system follow keplerian orbits
around the star and gives better results for a large number of
observational points.
The orbital parameters may present some errors, but they are an excellent
starting point for the traditional minimization methods such as the
Levenberg-Marquardt algorithms.
\end{abstract}

\section{Radial Velocity}
\label{sec:1}

The radial velocity of a star with $N$ companions is given by $ v (t) = \gamma +
v_0 (t) $, where $ \gamma $ is a drift due to the global shift of
the system center of mass, and\cite{Hilditch_2001}  
\begin{equation}
v_0 (t) = \sum_{j=1}^{N} K_j \left( e_j \cos \omega_j + \cos (\omega_j +
\nu_j ) \right) \ . \label{eq01}
\end{equation}
For each companion $j$, $ e_j $ is the eccentricity, $\omega_j $ is the argument of the pericentre, 
$ \nu_j = \nu_j (t) $ is the true anomaly and 
\begin{equation}
K_j = n_j a_j \, \frac{m_j}{\cal M} \, \sin I_j \, (1-e_j^2)^{-1/2} \ 
\label{eq02} 
\end{equation}
is the amplitude of the radial velocity variations. $ n_j $ is the mean motion, $ a_j
$ is the semi-major axis, $ I_j $ is the inclination of the orbital plane with
respect to the line of sight, $ m_j $ is the companion mass and $ {\cal M} $ is
the total mass of the system.
The orbital period of each companion is given by $ P_j = 2 \pi / n_j $.


There is no explicit expression for the true anomaly $ \nu_j (t) $, but making use
of the Kepler equation we can expand it in power series of $ e_j $ as\cite{Murray_Dermott_1999}:
\begin{equation}
\mathrm{e}^{i \nu_j} = \sum_{k=-\infty}^{+\infty} X_k (e_j) \, \mathrm{e}^{i k M_j} 
\ , \label{eq05} 
\end{equation}
where $ M_j = n_j (t-T_{0j} )$ is the mean anomaly, $T_{0j}$ is the date for the passage at the pericentre, and $X_k(e_j)$ are Hansen coefficients such that
\begin{equation}
X_k (e_j) 
= \frac{1}{2\pi} \int_{0}^{2\pi} \left( \cos E - e_j + i \sqrt{1-e_j^2}
\sin E \right) \mathrm{e}^{-i k (E - e_j \sin E)} \, d E \ . \label{eq06} 
\end{equation}
To the fifth order in eccentricity, $ X_k (e_j) $ for $ k=0, \pm 1, \pm 2$, becomes:
\begin{equation}
X_0 (e_j) = - e_j 
\ , \label{220609a} 
\end{equation}
\begin{equation}
X_1 (e_j) 
= 1 -  e_j^2 + \frac{7}{64} e_j^4 + \cdots 
 \ , \quad
X_{-1} (e_j) = - \frac18 e_j^2 + \frac{1}{48} e_j^4 + \cdots 
\ , \label{eq19} 
\end{equation}
\begin{equation}
X_2 (e_j) 
= e_j - \frac54 e_j^3 + \frac{17}{48} e^5 + \cdots 
 \ , \quad
X_{-2} (e_j) = - \frac{1}{12} e_j^3 + \frac{1}{48} e_j^5 + \cdots 
\ . \label{eq20} 
\end{equation}
Replacing expression (\ref{eq05}) in (\ref{eq01}), we rewrite the radial
velocity of the star as:
\begin{equation}
v_0 (t) =  \sum_{j=1}^{N} K_j \, \mathrm{e}^{i \omega_j} \sum_{k=-\infty}^{+\infty} C_k (e_j,\omega_j) \, \mathrm{e}^{- i k n_j T_{0j}} \, \mathrm{e}^{i k n_j t} 
\ , \label{eq07}  
\end{equation}
with $C_0 (e_j,\omega_j) = 0$, and for $k \ne 0$
\begin{equation}
C_k (e_j,\omega_j) = \frac12 \Big[ X_k (e_j)  + X_{-k} (e_j) \, \mathrm{e}^{-i 2 \omega_j} \Big] \ .
\end{equation}

\section{Fourier Analysis}
\label{sec:2}

For simplicity, we adopt an ordinary fast Fourier transform (FFT) of the radial velocity, 
\begin{equation}
F (\phi) \approx \frac{1}{T} \int_{0}^{T} v (t) \, \mathrm{e}^{-i \phi t} \, d t
\approx \frac{1}{T} \sum_{k=1}^{N_\mathrm{obs}} v_k \, \mathrm{e}^{-i \phi t_k}
(t_k-t_{k-1}) \ , \label{eq04} 
\end{equation}
where $ v_k $ is the star radial velocity measured for the date $ t_k $ and $ N_\mathrm{obs} $ is the discrete number of observations in a time span $ [0,T] $.
Note that more sophisticated FFT methods exist that ensure a better convergence with the observational data\cite{Zechmeister_Kuerster_2009, Delisle_etal_2016}, but the following analysis remains the same.

\subsection{Determination of $\gamma$}

Replacing $ v(t) = \gamma + v_0(t) $ in  expression (\ref{eq04}) with $ \phi = 0
$ we get: 
\begin{equation}
\gamma  = F(0) \ . \label{eq08} 
\end{equation}
It is then possible to have an estimation of $ \gamma $ using $ \phi = 0 $ in
expression (\ref{eq04}).
Once we have $ \gamma $ it is preferable to subtract its value from the
data $ v_k $ and then continue the Fourier analysis (Eq.\,\ref{eq04})
with the expression of $ v_0 (t) $ (Eq.\,\ref{eq07}).

\subsection{Determination of $n_j$}

The orbital frequency $ n_1 $ corresponding to the companion with largest
amplitude $ K_1$ is given by the frequency
$ \phi $ corresponding to the highest peak in the power spectrum, that is,
\begin{equation}
n_1 : \quad \forall \phi \ , \; | F(n_1) | \ge  | F (\phi) | \ .
\label{eq09}  
\end{equation}
After finding $ n_1 $ it is easy to determine the remaining orbital
parameters (see next section). 
Once the orbit of the first companion is completely established, it is recommended
to subtract its contribution from the data $ v_k $ and then continue the Fourier
analysis  (Eq.\,\ref{eq04}) with the expression of 
\begin{equation}
 v_0 - K_1 \left(e_1 \cos \omega_1 + \cos (\omega_1 + \nu_1) \right) \ .
\label{eq10}  
\end{equation}
We then have to repeat this procedure for all the other $ N - 1 $ remaining
companions of the star.
Thus, the $ n_j $ orbital frequencies are always given by the highest peak in the
spectrum (Eq.\ref{eq09}) after subtracting the signal from the already detected
companions (Eq.\ref{eq10}).

\subsection{Determination of the remaining orbital parameters}

Replacing expression (\ref{eq07}) in (\ref{eq04}) with $ \phi = n_j $ and $ \phi = 2 n_j $ we have
\begin{equation}
F(n_j) = K_j C_1 (e_j,\omega_j) \, \mathrm{e}^{i \omega_j}  \, \mathrm{e}^{-i n_j T_{0j}} 
\ , \label{eq11}  
\end{equation}
\begin{equation}
F(2 n_j) = K_j C_2 (e_j,\omega_j) \, \mathrm{e}^{i \omega_j}  \, \mathrm{e}^{-i 2 n_j T_{0j}} \ ,
\label{eq12}  
\end{equation}
where the quantities $ F(n_j) $ and $ F(2 n_j) $ can be computed from the data
using expression (\ref{eq04}).
We let
\begin{equation}
G(n_j) = F(2 n_j) \frac{| F(n_j) |}{F^2(n_j)} = C_2 (e_j,\omega_j) \frac{| C_1 (e_j,\omega_j)  |}{C_1^2 (e_j,\omega_j)}  \, \mathrm{e}^{- i \omega_j}  \ .
\end{equation}
The previous expression provides a system of two equations that allows us to determine $e_j$ and $\omega_j$ from the observations.
Denoting $\ez_j = e_j \, \mathrm{e}^{-i \omega_j}$, we have to the third order in eccentricity,
\begin{equation}
G(n_j) = \ez_j \left(1 
 + \frac{5}{48} \, \ez_j^2  -  \frac{1}{4} \, \ez_j \eb_j  -  \frac{1}{16} \, \eb_j^2 
 + \cdots \right) \ ,
\end{equation}that can be easily solved using 
an iteration method\cite{Press_etal_1992}.
As initial condition, we can simply adopt $\ez_j \approx G(n_j) $, 
which is already an excellent approximation for moderate eccentricities. 
Once we have determined the pair $(e_j,\omega_j)$, the remaining orbital parameters are obtained from expression (\ref{eq11}) as
\begin{equation}
K_j = \frac{| F(n_j) |}{| C_1 (e_j,\omega_j) |}  
\quad \mathrm{and} \quad 
\mathrm{e}^{-i n_j T_{0j}} = \frac{F(n_j) \, \mathrm{e}^{-i \omega_j}}{K_j C_1 (e_j,\omega_j) }   
\ . \label{eq15}  
\end{equation}
In appendix we provide the expressions of $G(z_j)$ and $ C_1 (e_j,\omega_j)$ up to $e_j^{14}$.

\section{Conclusion} 

For a single companion of a star, we are able to determine its orbital
parameters directly from the observational data by computing the FFTs for three
different frequencies, namely $ F(0) $, $ F(n) $ and $ F(2n) $.
We chose $ n $ and $ 2 n $, but according to expression (\ref{eq07}) we
could have chosen any frequency multiple of $ n $.
However, unless the eccentricity is extremely high, these two frequencies
correspond to the highest peaks produced by the companion in the spectrum and are
therefore easier to identify.
Moreover, if the eccentricity is close to zero 
(which is often the case for ``hot Jupiters'' and close binaries), 
$ F(kn) \approx 0 $, except for $ k = \pm 1 $. In this case, 
$ \omega_j $ and $ T_{0j} $ cannot be determined, but for a given time $t=t_0$, it is always possible
to establish the position of the companion in the orbit, $ \lambda_{j} = M_{j} + \omega_j $, as 
\begin{equation}
\mathrm{e}^{i \lambda_{j} } = \frac{F(n_j) \, \mathrm{e}^{i n_j t_0}}{K_j C_1 (e_j,\omega_j) } 
\ . \label{eq21}  
\end{equation}
The orbital parameters determined with our method present errors that are
proportional to the precision of the instrument and inversely proportional to the
number of data points, since a large number of points increases the convergence
with expression (\ref{eq04}).
The agreement between the Fourier parameters and the true parameters can be
increased if we perform a $ \chi^2 $ minimization after determining the orbit of
each companion. This procedure should be fast using a standard method such as 
a Levenberg-Marquardt algorithm\cite{Press_etal_1992}, since the Fourier
parameters are already close to the minimum value of $ \chi^2 $.
Even though the FFT method is established for keplerian orbits, it also works on
realistic systems for which planet-planet interactions are weak.
Indeed, this method has already been tested with success in the determination of
the orbital parameters of several different planetary
systems\cite{Correia_etal_2005, Lovis_etal_2006, Pepe_etal_2006, Delisle_etal_2016}, 
providing similar results as other classical alternative methods.


\vskip0.2truecm
{\small The author thanks Jean-Baptiste Delisle for discussions.}


\section*{Appendix}
Expressions of $G(e,\omega)$ and $ C_1 (e,\omega)$ up to $e^{14}$, with $\ez = e \, \mathrm{e}^{-i \omega}$ and $\eb = e \, \mathrm{e}^{i \omega}$.

\begin{eqnarray}
\small
\begin{split}
        G(\ez)/\ez
      & =    1
 + \frac{5}{48} \, \ez^2
 - \frac{1}{4} \, \ez \eb
 - \frac{1}{16} \, \eb^2 \\ &
  + \frac{7}{512} \, \ez^4 
 + \frac{3}{64} \, \ez^3 \eb
 - \frac{3}{256} \, \ez^2 \eb^2
 - \frac{7}{192} \, \ez \eb^3
 - \frac{1}{512} \, \eb^4 \\ &
 + \frac{15}{8192} \, \ez^6
 + \frac{103}{6144} \, \ez^5 \eb
 + \frac{1129}{40960} \, \ez^4 \eb^2
 - \frac{161}{9216} \, \ez^3 \eb^3 \\ & \quad 
 - \frac{641}{24576} \, \ez^2 \eb^4
 - \frac{17}{6144} \, \ez \eb^5
 - \frac{1}{8192} \, \eb^6 \\ &
  + 0.000244776407877604 \, \ez^8
 + 0.003712972005208333 \, \ez^7 \eb \\ & \quad 
+ 0.0158442179361979 \, \ez^6 \eb^2 
 + 0.01725147388599535 \, \ez^5 \eb^3 \\ & \quad 
 - 0.01402274237738737 \, \ez^4 \eb^4 
  - 0.0197475857204861 \, \ez^3 \eb^5 \\ & \quad 
- 0.002996656629774307 \, \ez^2 \eb^6 
 - 0.000274658203125 \, \ez \eb^7  \\ & \quad 
 - 9.5367431640625 \times 10^{-6} \, \eb^8 \\ &
 + 3.254413604736326 \times 10^{-5} \, \ez^{10}
 + 0.0006953875223795567 \, \ez^9 \eb  \\ & \quad 
 + 0.004983756277296273 \, \ez^8 \eb^2
+ 0.01377033657497829 \, \ez^7 \eb^3  \\ & \quad 
 + 0.01134570818098743 \, \ez^6 \eb^4 
  - 0.0117480511135527 \, \ez^5 \eb^5  \\ & \quad 
 - 0.01567484007941347 \, \ez^4 \eb^6
- 0.002966520521375871 \, \ez^3 \eb^7 \\ & \quad 
 - 0.0004082123438517253 \, \ez^2 \eb^8  
 - 2.940495808919271 \times 10^{-5} \, \ez \eb^9 \\ & \quad 
 - 8.344650268554688 \times 10^{-7} \, \eb^{10} \\ &
 + 4.302710294723508 \times 10^{-6} \, \ez^{12}
 + 0.0001191198825836181 \, \ez^{11} \eb \\ & \quad 
 + 0.001215890877776675 \, \ez^{10} \eb^2
 + 0.005628809884742449 \, \ez^9 \eb^3 \\ & \quad 
 + 0.01160760545935576 \, \ez^8 \eb^4
 + 0.007722734080420522 \, \ez^7 \eb^5 \\ & \quad 
 - 0.0100206465748466 \, \ez^6 \eb^6
 - 0.01287604926124456 \, \ez^5 \eb^7 \\ & \quad 
 - 0.002833366973532573 \, \ez^4 \eb^8
 - 0.0005092923287992127 \, \ez^3 \eb^9 \\ & \quad 
 - 5.576759576797487 \times 10^{-5} \, \ez^2 \eb^{10}
 - 3.268321355183919 \times 10^{-6} \, \ez \eb^{11} \\ & \quad 
 - 7.82310962677002 \times 10^{-8} \, \eb^{12} \\ &
\end{split}
\end{eqnarray}

\begin{eqnarray}
\small
\begin{split}
        C_1 (\ez)  
    &  =    \frac12
 - \frac12 \, \ez \eb
 - \frac{1}{16} \, \ez^2
 + \frac{7}{128} \, \ez^2 \eb^2
 + \frac{1}{96} \, \ez^3 \eb
 - \frac{5}{576} \, \ez^3 \eb^3
 + \frac{37}{6144} \, \ez^4 \eb^2 \\ &
 - 0.004594590928819439 \, \ez^4 \eb^4
 + 0.004649522569444453 \, \ez^5 \eb^3 \\ &
 - 0.00355889214409723 \, \ez^5 \eb^5
 + 0.00355820832429112 \, \ez^6 \eb^4 \\ &
 - 0.002806284633683508 \, \ez^6 \eb^6
 + 0.002806290318726159 \, \ez^7 \eb^5 \\ &
\end{split}
\end{eqnarray}

\end{document}